\documentclass[aps,floatfix,nofootinbib,onecolumn,superscriptaddress]{revtex4}
\pdfoutput=1

\usepackage{amsmath}
\usepackage{amssymb}
\usepackage{amsthm}
\usepackage{dcolumn}
\usepackage{epsfig}
\usepackage{graphics}
\usepackage{graphicx}
\usepackage{slashed,epsfig}
\usepackage{longtable}
\usepackage{color}

\definecolor{darkgreen}{rgb}{0,0.5,0}
\definecolor{purple}{rgb}{0.5,0,0.5}
\definecolor{nblue}{rgb}{0.0,0.0,0.50}
\definecolor{scarlet}{rgb}{1.0,0.2,0}
\usepackage[colorlinks=true, pdfstartview=FitV, linkcolor=purple, citecolor= purple, urlcolor=blue]{hyperref}

\newcommand{\beq} {\begin{equation}}
\newcommand{\eeq} {\end{equation}}
\newcommand{\beqa} {\begin{eqnarray}}
\newcommand{\eeqa} {\end{eqnarray}}

\begin{document}

{\par\raggedleft \texttt{SLAC-PUB-14704}\par}
\bigskip{}

\title{Transversity from First Principles in QCD}

\author{Stanley~J.~Brodsky} \affiliation{SLAC National Accelerator Laboratory\\
Stanford University, Stanford, California 94309, USA}

\begin{abstract}
Transversity observables, such as the T-odd single-spin asymmetry measured in deep inelastic lepton scattering on polarized protons,  and the distributions which are measured in deeply virtual Compton scattering provide important constraints on the fundamental  quark and gluon structure of the proton.  In this talk I discuss the challenge of computing these observables from first principles; i.e., quantum chromodynamics, itself.  A key step is the determination of 
the frame-independent light-front wavefunctions (LFWFs)  of hadrons -- the QCD eigensolutions which are analogs of the Schr\"odinger  wavefunctions of atomic physics. The lensing effects of initial-state and final-state interactions, acting on LFWFs with different orbital angular momentum, lead to the $T$-odd transversity observables such as the Sivers, Collins, and Boer-Mulders distributions.   
The lensing effect  also leads to leading-twist phenomena which break leading-twist factorization, such as the breakdown of the Lam-Tung relation in Drell-Yan reactions. A similar rescattering mechanism also leads to diffractive deep inelastic scattering,  as well as nuclear shadowing and non-universal antishadowing.  
It is thus important to distinguish ``static" structure functions, the probability distributions computed from the target hadron's light-front wavefunctions, versus ``dynamical" structure functions which include the effects of initial- and final-state rescattering.  I also discuss related effects, such as the $J=0$ fixed pole contribution which appears in the real part of the virtual Compton amplitude. AdS/QCD, together with ``Light-Front Holography", provides a simple Lorentz-invariant color-confining approximation to QCD which is successful in accounting for light-quark meson and baryon spectroscopy as well as hadronic LFWFs.
\end{abstract}


\maketitle

\date{\today}

\section{Introduction}
``Transversity" in hadron physics encompasses the entire range of spin, orbital angular momentum, and transverse momentum measures of hadron structure which are accessible by experiment.~\cite{Boer:2011fh,Lorce:2011zt,Barone:2010zz,Gamberg:2010xi,Anselmino:2011ay,Goldstein:2011af}  As we have seen at this meeting, highly sensitive experiments such as HERMES at DESY~\cite{Airapetian:2004tw}, COMPASS~\cite{Bradamante:2011xu,Alekseev:2010rw,Bradamante:2009zz} at CERN, and CLAS at Jefferson Laboratory~\cite{Avakian:2010ae,Gao:2010av} are now providing an extensive range of experimental results which are providing new insight into the fundamental quark and gluon structure of the nucleons. The challenge for theory is to synthesize this information into a consistent picture of hadron dynamics and to confront QCD at a fundamental level.

In the case of atomic physics the structure of atoms is described by  Schr\"odinger and Dirac wavefunctions.   In QCD, the corresponding relativistic, frame-independent bound-state amplitudes which describe the hadron's spin and momentum structure are the $n$-particle light-front Fock state wavefunctions $\Psi^H_n(x_i, \vec k_{\perp i}, \lambda_i) = <\Psi  \vert n>$  defined at fixed light-front time $\tau=t + z/c$.  
The constituents have light-front momentum fractions $x_i = k^+/P^+$ with $\sum^n_i  x_i =1,$ transverse momenta  $ \vec k_{\perp i}$  (with $ \sum^n_i \vec k_{\perp i} =0 $) parton spin-projections $\lambda_i$. 
Remarkably, the LFWFs $\Psi_n(x_i, \vec k_{\perp i,} \lambda_i)$  
are independent of the hadron's momentum $P^+=P^0 + P^z$ and $\vec P_\perp$, and thus they are independent of the observer's Lorentz frame.  The LFWFs are the coefficients of the expansion of a hadron eigenstate projected on a free Fock basis.  The sum includes the valence Fock state and higher Fock states with sea quark-antiquark pairs and gluons.
The total angular momentum $J^z= \sum_i^n  S^z _i+ \sum_i^{n-1}L^z_i$ is conserved in the LF formalism by every QCD interaction and within every Fock state,  just as in atomic physics. (Note that there are only $n-1 $ independent orbital angular momenta for an $n$-parton state. )  The light-front formalism thus provides a consistent, frame-independent definition of quark orbital angular momentum in hadrons.
 
The LFWF for a hadron $H$ is the eigensolution of the QCD light-front Hamiltonian satisfying the Heisenberg matrix equation $H^{LF}_{QCD}|\Psi_H > = M^2_H |\Psi_H>$ where $H_{QCD}$ 
is derived directly from the QCD Lagrangian~\cite{Brodsky:1997de}. The eigenvalues $M^2_H$ give the discrete and continuum hadron spectrum. If one chooses to quantize QCD in light-cone gauge $A^+=0$, the gluons have physical polarization $J^z= \pm 1$ and  ghost states  with negative norm are avoided. QCD(1+1) is solvable using matrix diagonalization for any number of colors, quark flavors and masses~\cite{Hornbostel:1988fb}, using the discretized light-front quantization method~\cite{Pauli:1985ps}. More generally, the LF Hamiltonian methods provide a frame-independent nonperturbative method for solving QCD(3+1)  in Minkowski space without fermion doubling.  The anomalous gravitomagnetic moment of each LF Fock state vanishes~\cite{Brodsky:2000ii}. as required by 
the equivalence theorem of gravity~\cite{Teryaev:1999su}.

\begin{figure}[t]
  \includegraphics[angle=0,width=1\textwidth]{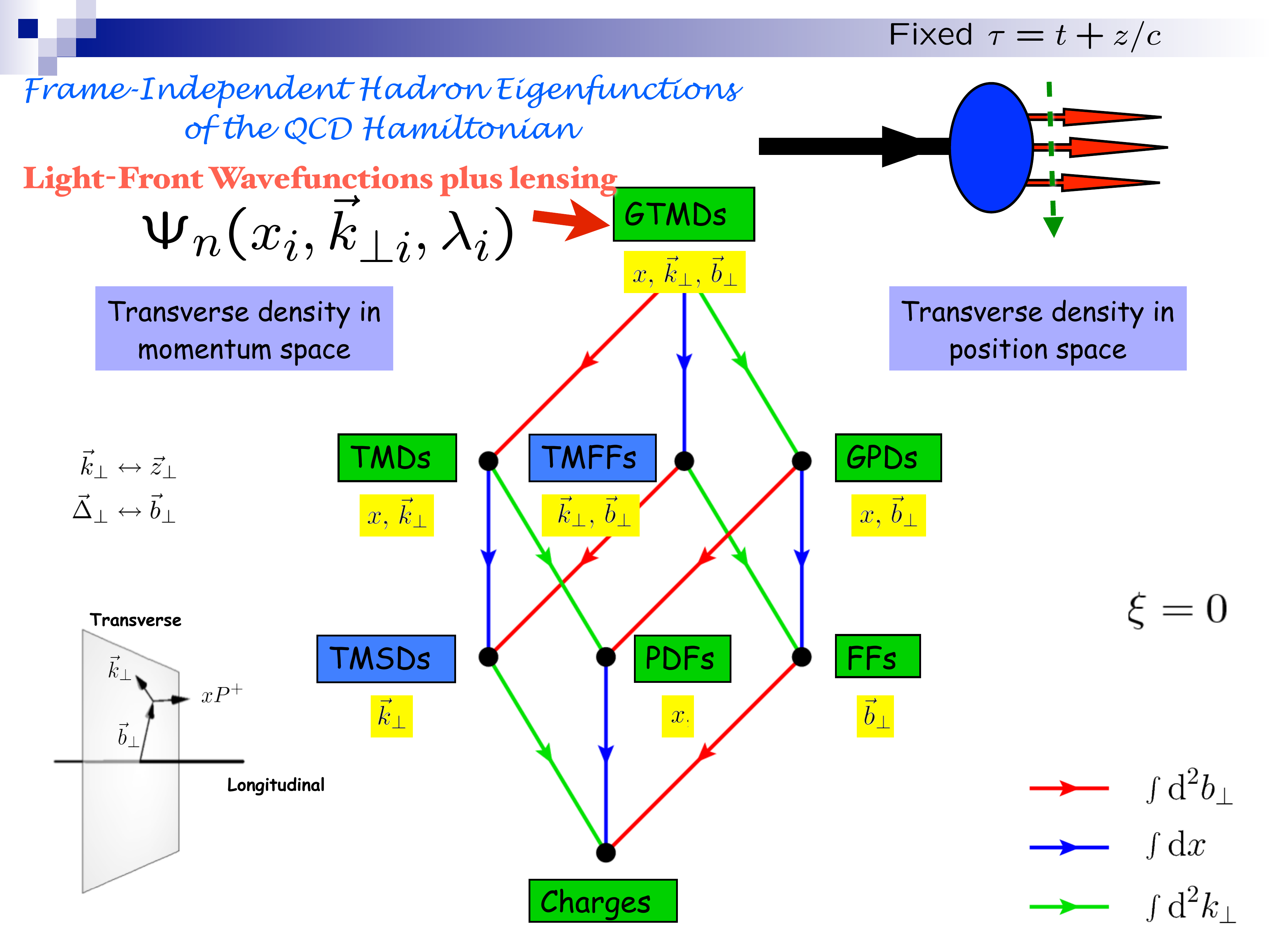}
  \caption{Transversity observables computable from first principles from hadron LFWFs plus lensing functions.  This figure was adapted from an illustration by F. Lorce'.}
\label{Lorce} 
\end{figure}

Given the frame-independent light-front
wavefunctions $\psi_{n/H}(x_i, \vec k_{\perp i}, \lambda_i )$, one can compute virtually all exclusive and inclusive hadron observables. 
See  fig.\ref{Lorce}.
For
example, the valence, sea-quark and gluon distributions are defined from the squares of the LFWFS summed over all Fock states $n$. Form factors,
exclusive weak transition amplitudes~\cite{Brodsky:1998hn} such as $B\to \ell \nu \pi,$ and the generalized parton
distributions~\cite{Brodsky:2000xy}, such as the distributions $E$ and $H$ measured in deeply virtual Compton scattering, are (assuming the ``handbag" approximation) overlaps of the
initial and final LFWFS with $n =n^\prime$ and $n =n^\prime+2$ (ERBL contributions).  
The hadron's distribution amplitude $\phi(x.Q)$ is the integral of the valence light front wavefunction in light-cone gauge integrated over transverse momenta $k^2_\perp < Q^2$.  ERBL and DGLAP evolution are automatically satisfied. 
Transversity observables can also be computed from the LFWFS; 
 However, in the case of pseudo-T-odd observables, one must include the lensing effect of final state or initial state interactions.

The LFWFs are derived from creation operators acting on the light-front vacuum, which unlike the usual ``instant form" vacuum, is causally-connected, frame-independent, and trivial.  Normal-ordering of the vacuum is thus not needed.  The LF  as well as the Bethe-Salpeter formalism thus predict zero cosmological constant~\cite{Brodsky:2009zd,Brodsky:2010xf}. 
In the front form, only constituents with positive $k^+$ occur. Thus form factors are simple overlaps of LFWFs. In contrast, in the instant form defined at fixed time $t$, one must include contributions to the current resulting from vacuum fluctuations such as the $q \bar q g$ currents which connect to the hadron, One must also boost the equal-time wavefunctions, a complicated dynamical problem~\cite{Brodsky:1968xc}.  

The five-quark Fock state of the proton's LFWF $|uud Q \bar Q>$ is the primary origin of the sea quark distributions of the proton~\cite{Brodsky:1996hc,Brodsky:2000sk}.  Experiments show that the sea quarks have remarkable nonperturbative features, such as $\bar u(x) \ne \bar d(x)$, an intrinsic strangeness~\cite{Airapetian:2008qf} distribution $s(x)$ appearing at $x > 0.1$, as well as intrinsic charm and bottom distributions at large $x$.  Such distributions~\cite{Brodsky:1984nx,Franz:2000ee} will arise rigorously from $g g \to Q \bar Q \to gg $ insertions connected to the valence quarks in the proton self-energy; in fact, they fit a universal intrinsic quark model~\cite{Brodsky:1980pb} as shown by Chang and Peng~\cite{Chang:2011du}. 

\section{AdS/QCD and Light-Front Holography}

A long-sought goal in hadron physics is to find a simple analytic first approximation to QCD analogous 
to the Schr\"odinger-Coulomb equation of atomic physics.	This problem is particularly challenging since the formalism must be relativistic, color-confining, and consistent with chiral symmetry.
de Teramond and I~\cite{deTeramond:2005su} have shown that
the soft-wall AdS/QCD model, modified by a positive-sign dilaton metric, leads to a simple 
Schr\"odinger-like light-front wave equation and a remarkable one-parameter description of nonperturbative hadron dynamics~\cite{deTeramond:2005su,deTeramond:2008ht,Brodsky:2010px,Brodsky:2011sk,arXiv:1105.3999,arXiv:1111.5169}. The model predicts a zero-mass pion for massless quarks and a Regge spectrum of linear trajectories with the same slope in the (leading) orbital angular momentum $L$ of the hadrons and their radial  quantum number $N$. 
 
Light-Front Holography maps the amplitudes which are functions of the fifth dimension variable $z$ of anti-de Sitter space to a corresponding hadron theory quantized on the light front. The resulting Lorentz-invariant relativistic light-front wave equations are functions of  an invariant impact variable $\zeta$ which measures the separation of the quark and gluonic constituents within the hadron at equal light-front time.  The result is 
a semi-classical frame-independent first approximation to the spectra and light-front wavefunctions of meson and baryon light-quark  bound states,  which in turn predicts  the
behavior of the pion and nucleon  form factors.  The theory implements chiral symmetry  in a novel way:    the effects of chiral symmetry breaking increase as one goes toward large interquark separation, consistent with spectroscopic data,   
The hadron eigenstates generally have components with different orbital angular momentum; e.g.,  the proton eigenstate in AdS/QCD with massless quarks has $L^z=0$ and $L^z=1$ light-front Fock components with equal probability -- the proton acts as a chiral dual.    Thus in AdS/QCD the spin of the proton is carried by the quark orbital angular momentum: $J^z=  <L^z>=\pm 1/2$ since $\sum S^z_q = 0,$  helping to explain the ``spin-crisis"~\cite{Valery}

\begin{figure}
 \begin{center}
\includegraphics[width=18cm]{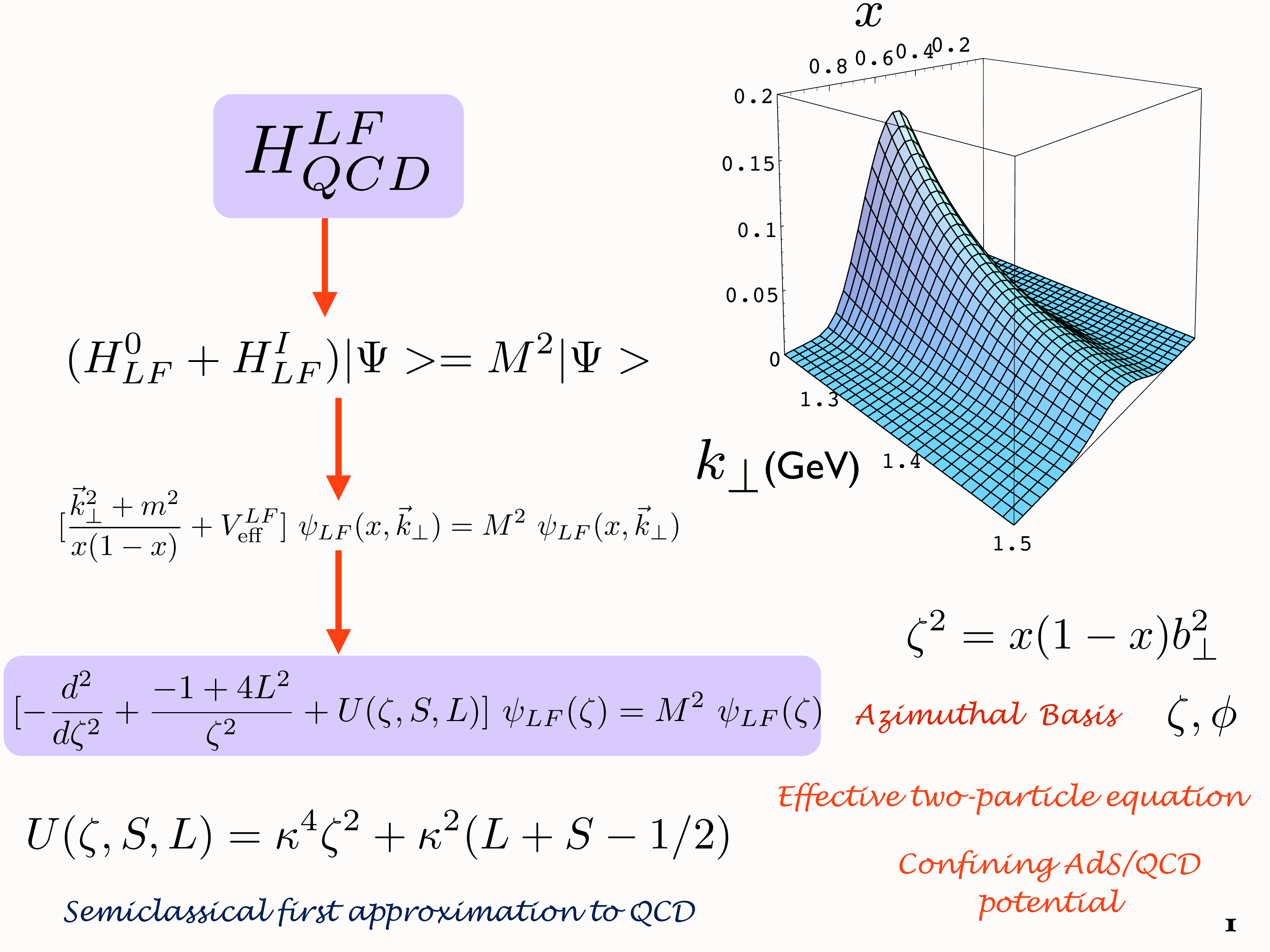}
\end{center}
\caption{Reduction of the Light-Front Hamiltonian to an effective LF Schrodinger Equation for mesons.  The insert shows the AdS/QCD -- light-front holography prediction for the pion's valence LFWF $\psi(x, \vec k_\perp).$ 
From ref.~\cite{deTeramond:2005su}.}
\label{HLF}  
\end{figure}

The AdS/QCD soft-wall model also predicts the form of the non-perturbative effective coupling $\alpha_s^{AdS}(Q)$ and its $\beta$-function~\cite{Brodsky:2010ur}, and  the AdS/QCD light-front wavefunctions also lead to a method for computing the hadronization of quark and gluon jets at the amplitude level~\cite{Brodsky:2008tk}.
In general the QCD Hamiltonian can be systematically reduced to an effective equation in  acting on the valence Fock state.
This is illustrated for mesons in fig.~\ref{HLF} 
The kinetic energy contains a term ${-1+4 L^2/ \zeta^2}$ analogous to $\ell(\ell+1)/r^2$ in nonrelativistic theory, where the invariant $\zeta^2 = x(1-x)b^2_\perp$  is conjugate to the $q \bar q $ invariant mass  $k^2_\perp/x(1-x)$. It  plays the role of the radial variable $r$.  Here $L= L^z$ is the projection of the orbital angular momentum appearing in the $\zeta, \phi$ basis. In QCD, the interaction $U$ couples  the valence state to all Fock states.  The AdS/QCD model has the identical structure as the reduced form of the LF Hamiltonian, but it also specifies the confining potential as $U(\zeta,S,L) = \kappa^4\zeta^2+ \kappa^2(L + S - 1/2).$ This correspondence, plus the fact that one can match the AdS/QCD formulae for elastic electromagnetic and gravitational form factors to the LF Drell-Yan West formula, is the basis for light-front holography.  The light-quark meson and baryon spectroscopy is  well described taking the mass parameter $\kappa \simeq 0.5$ GeV.  The linear trajectories in $M^2_H(n,L)$  have the same slope in $L$ and $n$, the radial quantum number.  The corresponding LF wavefunctions are functions of the off-shell invariant mass.   AdS/QCD, together with Light-Front Holography~\cite{deTeramond:2005su} thus provides an simple Lorentz-invariant color-confining approximation to QCD which is successful in accounting for light-quark meson and baryon spectroscopy as well as their LFWFs.  It can be systematically improved by Lippmann-Schwinger methods~\cite{Hiller} or using the AdS/QCD orthonormal basis to diagonalize $H^{QCD}_{LF}$ ~\cite{Vary}.

\section{The Real Part of the DVCS amplitude}

It is usually assumed that the imaginary part of the deeply virtual Compton scattering amplitude is determined at leading twist by  generalized parton distributions, but that the real part has an undetermined  ``$D$-term" subtraction. In fact, the real part is determined by the  local  two-photon interactions of the quark current in the QCD light-front Hamiltonian~\cite{Brodsky:2008qu,Brodsky:1971zh}.  This contact interaction leads to a real energy-independent contribution to the DVCS amplitude  which is independent of the photon virtuality at fixed  $t$.  The interference of the timelike DVCS amplitude with the Bethe-Heitler amplitude leads to a charge asymmetry in $\gamma p \to \ell^+ \ell^- p$~\cite{Brodsky:1971zh,Brodsky:1973hm,Brodsky:1972vv}.    Such measurements can verify that quarks carry the fundamental electromagnetic current within hadrons.

\section{Lensing and the Sivers Effect}

The effects of final-state interactions of the scattered quark  in deep inelastic scattering  have been traditionally assumed to be power-law suppressed.  In fact,  the final-state gluonic interactions of the scattered quark lead to a  $T$-odd non-zero spin correlation of the plane of the lepton-quark scattering plane with the polarization of the target proton~\cite{Brodsky:2002cx}.  This  leading-twist Bjorken-scaling ``Sivers effect"  is nonuniversal since QCD predicts an opposite-sign correlation~\cite{Collins:2002kn,Brodsky:2002rv} in Drell-Yan reactions due to the initial-state interactions of the annihilating antiquark. 
The same final-state interactions of the struck quark with the spectators~\cite{Brodsky:2002ue}  also lead to diffractive events in deep inelastic scattering (DDIS) at leading twist,  such as $\ell p \to \ell^\prime p^\prime X ,$ where the proton remains intact and isolated in rapidity;    in fact, approximately 10\% of the deep inelastic lepton-proton scattering events observed at HERA are
diffractive~\cite{Adloff:1997sc,Breitweg:1998gc}.  The presence of a rapidity gap
between the target and diffractive system requires that the target
remnant emerges in a color-singlet state; this is made possible in
any gauge by the soft rescattering incorporated in the Wilson line or by augmented light-front wavefunctions~\cite{Brodsky:2010vs}.

\begin{figure}
 \begin{center}
\includegraphics[width=15cm]{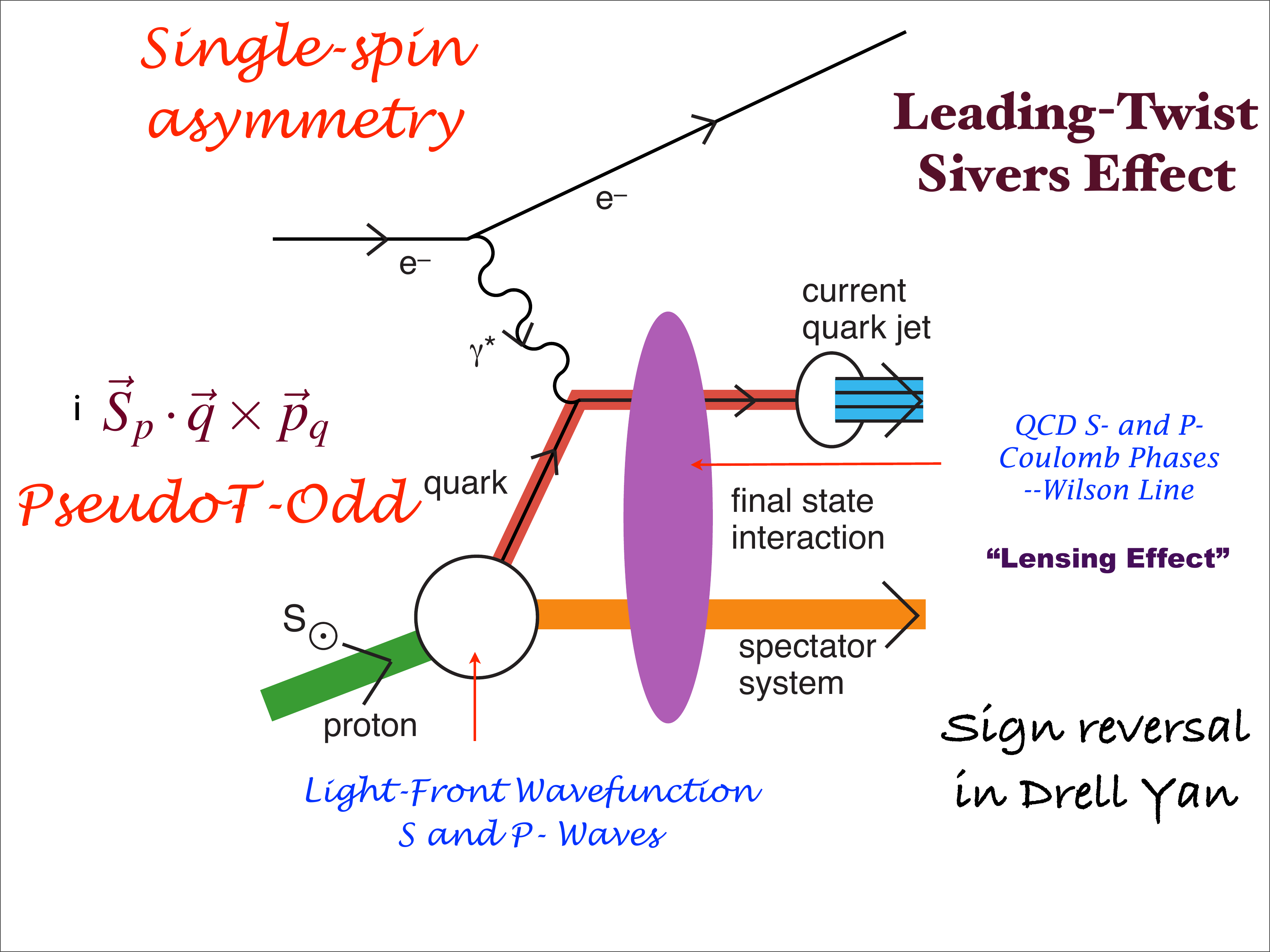}
\end{center}
\caption{Origin of the Sivers single-spin asymmetry in deep inelastic lepton scattering.}
\label{Sivers}  
\end{figure}

\begin{figure}
 \begin{center}
\includegraphics[width=19cm]{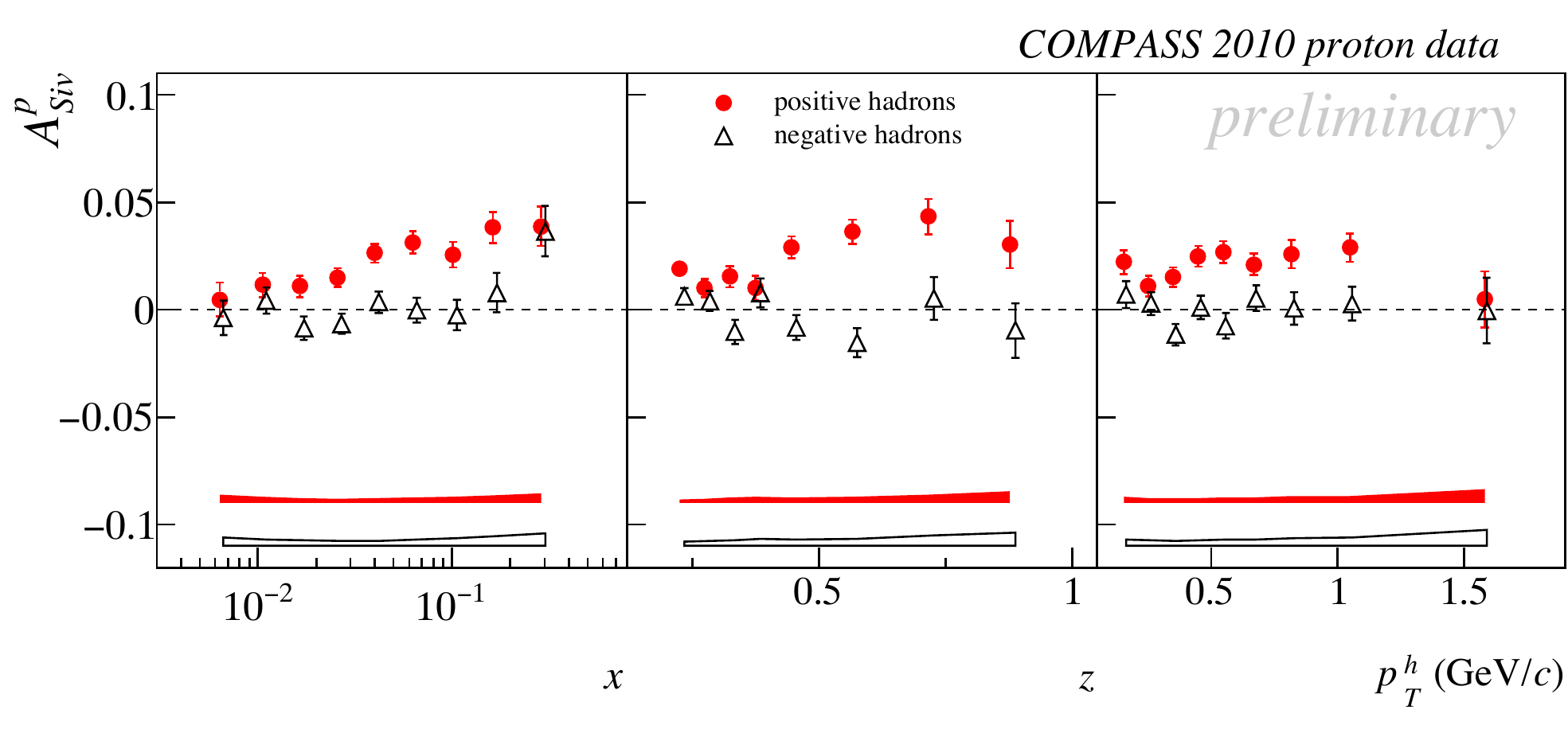}
\end{center}
\caption{Preliminary measurements of the Sivers single-spin asymmetry by COMPASS in deep inelastic muon-proton scattering.}
\label{Compass}  
\end{figure}

The calculation of the Sivers single-spin asymmetry in deep inelastic lepton scattering is illustrated in fig.~\ref{Sivers}.
The analysis requires two different orbital angular momentum components: $S$-wave with the quark-spin parallel to the proton spin and $P$-wave for the quark with anti-parallel spin; the difference between the final-state ``Coulomb" phases leads to the $\vec S \cdot \vec q \times \vec p$ correlation of the proton's spin with the virtual photon-to-quark production plane.  Recent high precision measurements presented at Transversity 2011 by the COMPASS collaboration  from $\mu p \to \mu^\prime H^\pm X$ is shown in fig.~\ref{Compass}.
The original model calculation by Hwang, Schmidt, and myself~\cite{Brodsky:2002ue} in fact gives a good representation of the HERMES and COMPASS data.
The same $S-$ and $P$-wave proton wavefunctions appear in the calculation of the Pauli form factor quark-by-quark. Thus one can correlate the Sivers asymmetry for each struck with the anomalous magnetic moment of the proton carried by that quark~\cite{Lu:2006kt}, thus leading to the prediction that the Sivers effect is larger for  positive pions.
The empirical evidence that the Sivers effect is small for the deuteron suggests that gluons do not carry significant orbital angular momentum in the nucleon.

One also can associate the dynamics of lensing with a Wilson line for each partial wave~\cite{Brodsky:2010vs}.  The physics of the lensing dynamics involves nonperturbative quark-quark interactions at small momentum transfer, not  the hard scale $Q^2$  of the virtuality of the photon.  It would interesting to see if these strength soft initial or final state scattering interactions can be predicted using the confining potential of AdS/QCD.

\section{Dynamic versus Static Hadronic Structure Functions}
The nontrivial effects from rescattering and diffraction highlight the need for a fundamental understanding the dynamics of hadrons in QCD at the amplitude
level. This is essential for understanding phenomena such as the quantum mechanics of hadron formation, the remarkable
effects of initial and final interactions, the origins of diffractive phenomena and single-spin asymmetries, and manifestations of higher-twist
semi-exclusive hadron subprocesses. 

It is natural to assume that the nuclear modifications to the structure functions measured in deep inelastic lepton-nucleus and neutrino-nucleus interactions are identical;  in fact,  Gribov-Glauber theory predicts that the antishadowing of nuclear structure functions is not  universal, but depends on the quantum numbers of each struck quark and antiquark~\cite{Brodsky:2004qa}.   This observation can explain the recent analysis of Schienbein et al.\cite{Schienbein:2008ay} which shows that the NuTeV measurements of nuclear structure functions obtain from neutrino  charged current reactions differ significantly from the distributions measured in deep inelastic electron and muon scattering.
  
As noted by Collins and Qiu~\cite{Collins:2007nk}, the traditional factorization formalism of perturbative QCD  fails in detail for many types of hard inclusive reactions because of initial- and final-state interactions.  For example, if both the
quark and antiquark in the Drell-Yan subprocess
$q \bar q \to  \mu^+ \mu^-$ interact with the spectators of the
other  hadron, then one predicts a $\cos 2\phi \sin^2 \theta$ planar correlation in unpolarized Drell-Yan
reactions~\cite{Boer:2002ju}.   This ``double Boer-Mulders effect" can account for the large $\cos 2 \phi$ correlation and the corresponding violation~\cite{Boer:1999mm,Boer:2002ju} of the Lam-Tung relation for Drell-Yan processes observed by the NA10 collaboration.   
An important signal for factorization breakdown at the LHC  will be the observation of a $\cos 2 \phi$ planar correlation in dijet production.

\begin{figure}
 \begin{center}
\includegraphics[width=14cm]{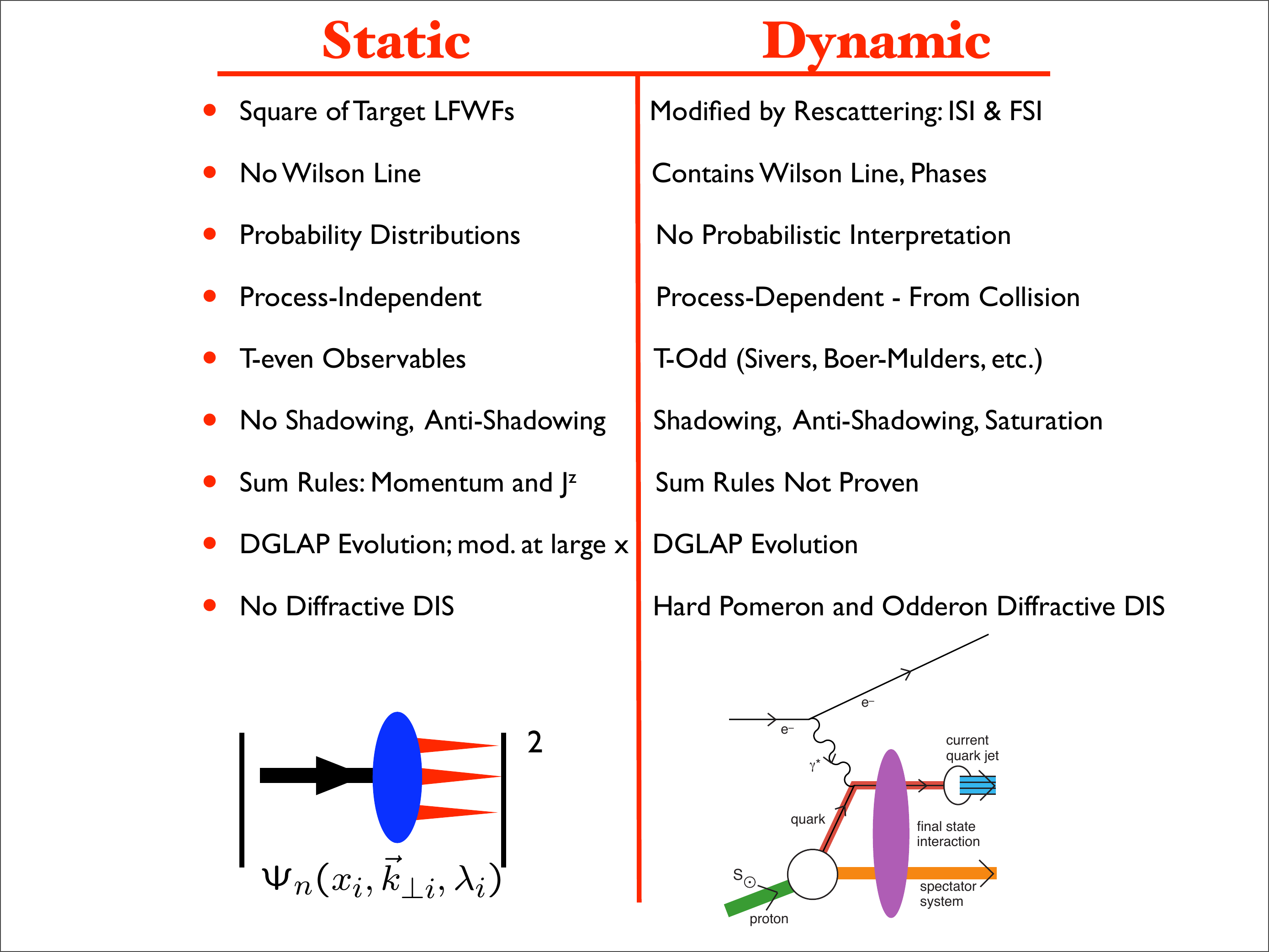}
\end{center}
\caption{Static versus dynamic structure functions}
\label{figNew17}  
\end{figure}

It is thus important to distinguish~\cite{Brodsky:2009dv} ``static" structure functions which are computed directly from the light-front wavefunctions of  a target hadron from the nonuniversal ``dynamic" empirical structure functions which take into account rescattering of the struck quark in deep inelastic lepton scattering. 
See  fig.\ref{figNew17}.
The real wavefunctions of hadrons which underlying the static structure functions cannot describe diffractive deep inelastic scattering nor  single-spin asymmetries, since such phenomena involve the complex phase structure of the $\gamma^* p $ amplitude.   
One can augment the light-front wavefunctions with a gauge link corresponding to an external field
created by the virtual photon $q \bar q$ pair
current~\cite{Belitsky:2002sm,Collins:2004nx}, but such a gauge link is
process dependent~\cite{Collins:2002kn}, so the resulting augmented
wavefunctions are not universal.
\cite{Brodsky:2002ue,Belitsky:2002sm,Collins:2003fm}.  The physics of rescattering and nuclear shadowing is not
included in the nuclear light-front wavefunctions, and a
probabilistic interpretation of the nuclear DIS cross section is thus
precluded. 

\section{Transversity in Hadron-Hadron Scattering}

A historic example of transversity is the remarkably large  spin correlation $A_{NN}$ measured in elastic $p p$ elastic scattering~\cite{Court:1986dh} measured by Krisch and collaborators, where the beam and target are polarized normal to the scattering plane. Remarkably the ratio of spin parallel to anti-parallel scattering reaches 4:1 at $\sqrt s \simeq 5 GeV.$ This can be explained~\cite{Brodsky:1987xw} as due to the excitation of a $uud uud c \bar c$ resonance
with $J=L=S=1$ in the intermediate state.  A comparable effect is also seen at the $\phi$ threshold

\section{Conclusions}
Transversity observables, such as the T-odd Sivers effect measured in deep inelastic lepton scattering on polarized protons, and the momentum distributions measured in deeply virtual Compton scattering are now providing new constraints on the fundamental  quark and gluon structure of the proton. In this talk I have discussed the challenge of computing these observables from first principles; i.e.; quantum chromodynamics, itself.  A key step is the determination of the 
the frame-independent light-front wavefunctions (LFWFs)  of hadrons -- the analogs of the Schr\"odinger  wavefunctions of atomic physics. They are  the eigenstates of the QCD Hamiltonian evaluated at fixed light-front time and the fundamental amplitudes underlying hadron observables such as transverse momentum distributions, structure functions and distribution amplitudes. 

The lensing effects of initial-state and final-state interactions together with the LFWFs with different orbital angular momentum lead to $T$-odd transversity observables such as the Sivers, Collins, and Boer-Mulders distributions.   
The lensing effect  also leads to leading-twist phenomena which break factorization theorems, such as the breakdown of
the Lam-Tung relation in Drell-Yan reactions. A similar rescattering mechanism also leads to diffractive deep inelastic scattering,  as well as nuclear shadowing and antishadowing.  
Thus one should distinguish ``static" structure functions, the probability distributions computed from the square of the target light-front wavefunctions, versus ``dynamical" structure functions which include the effects of initial- and final-state rescattering.  I have also discussed related effects such as the $J=0$ fixed pole contribution which appears in the real part of the virtual Compton amplitude.   Remarkably, AdS/QCD, together with ``Light-Front Holography" provides an simple Lorentz-invariant color-confining approximation to QCD which is successful in accounting for light-quark meson and baryon spectroscopy as well as the QCD dynamics expressible in terms of their LFWFs.

\begin{acknowledgements}
Presented at Transversity 2011: The 3rd International Workshop On Transverse Polarization Phenomena In Hard Scattering 29 August - 2 September 2011, Veli Lozinj, Croatia.
I am grateful to Anna Martin and Franco Bradamante for their invitation to Transversity 2011,  and I thank all of my collaborators whose work has been cited in this report, particularly Guy de Teramond, Ivan Schmidt, and Dae Sung Hwang. This research was supported by the Department of Energy,  contract DE--AC02--76SF00515. 
\end{acknowledgements}

\end{document}